\documentclass[a4paper,10pt]{article}
\usepackage[english]{babel}
\usepackage[utf8]{inputenc}

\usepackage[nottoc]{tocbibind}
\usepackage{amsmath}
\usepackage{amssymb}
\usepackage{mathtools}
\usepackage{bm}
\usepackage{blkarray, bigstrut} %
\usepackage{gauss}
\usepackage{graphicx}
\usepackage{float}
\usepackage{subfig}
\usepackage{booktabs}
\usepackage[table,xcdraw]{xcolor}
\usepackage{makecell}
\usepackage{multirow}
\usepackage{amsfonts,amsthm,epsfig,epstopdf,titling,url,array}
\usepackage{color, colortbl}
\usepackage{array}
\usepackage{diagbox}
\usepackage[thinlines]{easytable}
\usepackage{tabu}

\usepackage{nicematrix}

\theoremstyle{plain}
\newtheorem{theorem}{Theorem}[section]

\theoremstyle{definition}

\theoremstyle{remark}

\usepackage{hyperref}
\hypersetup{
    colorlinks=true,
    linkcolor=blue,
    filecolor=blue,      
    urlcolor=blue,
}

\usepackage{tikz}
\usetikzlibrary{shapes,backgrounds}

\usepackage{indentfirst}
 
\title{Evolutionary Dynamics with Randomly Distributed Benevolent Individuals}
\author
{Yuxin Geng,${}^{1,2}$ and Xingru Chen${}^{1,2}$\\
${}^{1}$Beijing University of Posts and Telecommunications, \\ Beijing, 100876, China\\
${}^{2}$Key Laboratory of Mathematics and Information Networks \\(Beijing University of Posts and Telecommunications), \\ Ministry of Education, China
}

\begin{document}

\maketitle

\newpage

\begin{abstract}
    Understanding the evolution of cooperation is pivotal in biology and social science. Public resources sharing is a common scenario in the real world. In our study, we explore the evolutionary dynamics of cooperation on a regular graph with degree $k$, introducing the presence of a third strategy, namely the benevolence, who does not evolve over time, but provides a fixed benefit to all its neighbors. We find that the presence of the benevolence can foster the development of cooperative behavior and it follows a simple rule: $b/c > k - p_S(k-1)$. Our results provide new insights into the evolution of cooperation in structured populations.
\end{abstract}

\begin{center}
\end{center}

\vspace{5ex}

\tableofcontents

\medskip

\newpage

\section{Introduction}


Cooperation requires an individual to bear a cost and sacrifice their own well-being to confer a benefit upon others. A paradigm to study cooperation is the Prisoner's dilemma in game theory\cite{axelrod1981evolution}. In this game, two players can choose to cooperate or defect. If both players cooperate, they both receive a reward ($R$). If one player cooperates while the other defects, the cooperating player receives the sucker's payoff ($S$), and the defecting player receives the temptation payoff ($T$). If both players choose to defect, they both get the punishment payoff ($P$). The four payoffs satisfy the following inequalities: $T > R > P > S$, and $2R > T + S$.

Whatever the chosen strategy of the other individual, defection is always the best response, while two cooperators can achieve a higher payoff, which leads to the so-called social dilemma\cite{macy2002learning}. On the other hand, cooperation is ubiquitous in nature, and the question of how cooperation can be established and maintained has attracted much attention\cite{nowak2010evolution}. 

Evolutionary game theory provides a mathematical framework for studying how the behaviors of the population evolve\cite{smith1973logic,smith1982evolution}. In this context, individuals are assumed to be selfish and rationality-bounded players, who interact with each other, get payoffs or fitness value from the given game, and adjust their strategies accordingly. Based on this framework, many mechanisms have been proposed to explain the evolution of cooperation, including network reciprocity\cite{nowak2006five}. In this setting, individuals are not well-mixed, but are connected by a network, and only interact with their neighbors\cite{lieberman2005evolutionary}.

Previous studies have shown that cooperation can be favored by weak selection in structured populations, if the benefit-to-cost ratio $b/c$ is greater than the average degree of the network $k$ \cite{ohtsuki2006simple}. However, real-life decision-making scenarios are far more complex than this simplified binary model. For instance, the sharing of resources in internet-based communities presents a scenario where the traditional binary strategies may not fully capture the range of interactions and outcomes. In our study, we introduce a third strategy to the system, who provides a fixed benefit to all other individuals, and does not evolve over time. We call this strategy the benevolence. Through theoretical analysis and simulations, we demonstrate that under certain conditions, the introduction of these strategists can foster the evolution of cooperative behavior. This study offers new perspectives for understanding and designing more complex evolutionary game models, holding theoretical and practical implications for explaining cooperative behaviors in the real world.


\section{Model}

Consider the payoff matrix:
\begin{equation*}
    \begin{pNiceMatrix}[first-row,first-col]
        & A & B & S \\
        A & a & b & \eta \\
        B & c & d & \eta \\
        S & 0 & 0 & 0 \\
    \end{pNiceMatrix}
\end{equation*}

We assume a population of $N$ individuals is distributed on a regular graph of degree $k$. Each individual can be $A$-player, $B$-player or $S$-player.

We introduce the following notations:
\begin{equation*}
    \begin{aligned}
    p_X &: \text{the frequency of $X$ players in the population}, \\
    p_{XY} &: \text{the frequency of $XY$ pairs in the population}, \\
    q_{X|Y} &: \text{the conditional probability to find an $X$ player given that the adjacent player is $Y$}. \\
    \end{aligned}
\end{equation*}

We assume that the the $S$-players are randomly distributed in the population, and they do not evolve their strategies. Therefore, we have the following equations:
\begin{equation*}
    \begin{aligned}
        p_A + p_B + p_S &= 1 \\
        q_{A|X} + q_{B|X} + q_{S|X} &= 1, \\
        p_{XY} &= q_{X|Y} \cdot p_Y, \\
        p_{XY} &= p_{YX},
    \end{aligned}
\end{equation*}
where $X, Y \in \{A, B, S\}$.

From our assumptions about the $S$-players, we have
\begin{equation*}
    \begin{aligned}
        p_S \text{ is fixed}, \\
        q_{S|X} = p_S.
    \end{aligned}
\end{equation*}

The whole system can be fully described by $p_A$, $q_{A|A}$, and $p_S$.

We use the death-birth (DB) updating rule. In each time step, an individual is chosen with uniform probability to be eliminated. Then one of its neighbors is chosen with probability proportional to their fitness to reproduce. The offspring replaces the vacant site. 

\section{Derivation}

\subsection{Updating a $B$-player}

A $B$-player is chosen to be eliminated. The probability that it has $k_A$, $k_B$ and $k_S$ many of $A$, $B$ and $T$ neighbors respectively is given by the multinomial distribution:
\begin{equation*}
    \frac{k!}{k_A! k_B! k_S!} q_{A|B}^{k_A} q_{B|B}^{k_B} q_{S|B}^{k_S}
\end{equation*}
The fitness of each $A$-player is given by:
\begin{equation*}
    f_A = 1 + \omega \cdot \left\{-1 + (k-1)q_{A|A} \cdot a + \left[(k-1)q_{B|A} + 1\right] \cdot b + (k-1)q_{S|A} \cdot \eta \right\}
\end{equation*}
The fitness of each $B$-player is given by:
\begin{equation*}
    f_B = 1 + \omega \cdot \left\{-1 + (k-1)q_{A|B} \cdot c + \left[(k-1)q_{B|B} + 1\right] \cdot d + (k-1)q_{S|B} \cdot \eta \right\}
\end{equation*}
The fitness of each $S$-player is given by:
\begin{equation*}
    f_S = 1 
\end{equation*}
The probability that one of the $A$-players is chosen to reproduce is given by:
\begin{equation*}
    \frac{k_Af_A}{k_A f_A + k_B f_B + k_S f_S}
\end{equation*}
The probability that $p_A$ increases by $1/N$ is given by:
\begin{equation*}
    \mathbf{P}(\Delta p_A = \frac1N) = p_B \cdot \sum_{k_A+k_B+k_S = k}\frac{k!}{k_A! k_B! k_S!} q_{A|B}^{k_A} q_{B|B}^{k_B} q_{S|B}^{k_S} \cdot \frac{k_Af_A}{k_A f_A + k_B f_B + k_S f_S}
\end{equation*}
The probability that $p_{AA}$ increases by $k_A/(kN/2)$ is given by:
\begin{equation*}
    \mathbf{P}(\Delta p_{AA} = \frac{k_A}{kN/2}) = p_B \cdot \frac{k!}{k_A! k_B! k_S!} q_{A|B}^{k_A} q_{B|B}^{k_B} q_{S|B}^{k_S}  \cdot \frac{k_Af_A}{k_A f_A + k_B f_B + k_S f_S}
\end{equation*}

\subsection{Updating a $A$-player}

A $A$-player is chosen to be eliminated. The probability that it has $k_A$, $k_B$ and $k_S$ many of $A$, $B$ and $T$ neighbors respectively is given by the multinomial distribution:
\begin{equation*}
    \frac{k!}{k_A! k_B! k_S!} q_{A|A}^{k_A} q_{B|A}^{k_B} q_{S|A}^{k_S}
\end{equation*}

The fitness of each $A$-player is given by:
\begin{equation*}
    g_A = 1 + \omega \cdot \left\{-1 + \left[(k-1)q_{A|A} + 1\right] \cdot a + (k-1)q_{B|A}  \cdot b + (k-1)q_{S|A} \cdot \eta \right\}
\end{equation*}

The fitness of each $B$-player is given by:
\begin{equation*}
    g_B = 1 + \omega \cdot \left\{-1 + \left[(k-1)q_{A|A} + 1\right] \cdot c + (k-1)q_{B|A}  \cdot d + (k-1)q_{S|B} \cdot \eta \right\}
\end{equation*}

The fitness of each $S$-player is given by:
\begin{equation*}
    g_S = 1
\end{equation*}

The probability that one of the $B$-players is chosen to reproduce is given by:
\begin{equation*}
    \frac{k_Bg_B}{k_A g_A + k_B g_B + k_S g_S}
\end{equation*}

The probability that $p_A$ decreases by $1/N$ is given by:
\begin{equation*}
    \mathbf{P}(\Delta p_A = -\frac1N) = p_A \cdot \sum_{k_A+k_B+k_S = k}\frac{k!}{k_A! k_B! k_S!} q_{A|A}^{k_A} q_{B|A}^{k_B} q_{S|A}^{k_S} \cdot \frac{k_Bg_B}{k_A g_A + k_B g_B + k_S g_S}
\end{equation*}
The probability that $p_{AA}$ decreases by $k_A/(kN/2)$ is given by:
\begin{equation*}
    \mathbf{P}(\Delta p_{AA} = -\frac{k_A}{kN/2}) = p_A \cdot \frac{k!}{k_A! k_B! k_S!} q_{A|A}^{k_A} q_{B|A}^{k_B} q_{S|A}^{k_S}  \cdot \frac{k_Bg_B}{k_A g_A + k_B g_B + k_S g_S}
\end{equation*}

\subsection{Different time scales}

We assume that one replacement occurs in each time step. Then the time derivative of $p_A$ and $p_{AA}$ are given by:
\begin{equation*}
    \begin{aligned}
        \dot{p}_A &= \frac{1}{N} \mathbf{P}(\Delta p_A = \frac1N) - \frac{1}{N} \mathbf{P}(\Delta p_A = -\frac1N) \\
        &= \omega \frac{k-1}{Nk} \cdot p_{AB} \cdot \left[(F_A-G_B) - q_{A|B}F_A - q_{B|B}F_B - q_{S|B}F_S + q_{B|A}G_B + q_{A|A}G_A + q_{S|A}G_S\right] + O(\omega^2)
    \end{aligned}
\end{equation*}
and
\begin{equation*}
    \begin{aligned}
        \dot{p}_{AA} &= \sum_{k_A+k_B+k_S = k}\frac{2k_A}{kN} \cdot \mathbf{P}\left(\Delta p_{AA}= \frac{k_A}{kN/2}\right) + \sum_{k_A+k_B+k_S = k}\left(-\frac{2k_A}{kN}\right) \cdot \mathbf{P}\left(\Delta p_{AA}= -\frac{k_A}{kN/2}\right)\\
        &= \frac{2}{kN} \cdot p_{AB} \cdot \left[1 + (k-1)(q_{A|B} - q_{A|A})\right] + O(\omega), \\
    \end{aligned}
\end{equation*}
where
\begin{equation*}
    \begin{aligned}
        F_A = \frac{\mathrm{d}f_A}{\mathrm{d}\omega}, \quad F_B = \frac{\mathrm{d}f_B}{\mathrm{d}\omega}, \quad f_S = \frac{\mathrm{d}f_S}{\mathrm{d}\omega}, \\
        G_A = \frac{\mathrm{d}g_A}{\mathrm{d}\omega}, \quad G_B = \frac{\mathrm{d}g_B}{\mathrm{d}\omega}, \quad g_S = \frac{\mathrm{d}g_S}{\mathrm{d}\omega}.
    \end{aligned}
\end{equation*}

\vspace{3ex}

Thus $q_{A|A}$ can be obtained by
\begin{equation*}
    \begin{aligned}
        \dot{q}_{A|A} = \frac{\mathrm{d}}{\mathrm{d}t} \left(\frac{p_{AA}}{p_A}\right) = \frac{\dot{p}_{AA}p_A - p_{AA}\dot{p}_A}{p_A^2}
    \end{aligned}
\end{equation*}

When $\omega \to 0$, $q_{A|A}$ equilibrates much faster than $p_A$. Therefore, we can set $\dot{q}_{A|A} = 0$, which leads to $\dot{p}_{AA} = 0$. Then we have the following equation:
\begin{equation} \label{eq:relation}
    q_{A|A} - q_{A|B} = \frac{1}{k-1}
\end{equation}
or
\begin{equation*}
    q_{B|B} - q_{B|A} = \frac{1}{k-1}
\end{equation*}

Note that this relation is the same as the one obtained in the case of no benevolence\cite{ohtsuki2006simple}.

In the following, we assume that the above equations always holds.

\subsection{Fixation probability}

We now calculate the fixation probability of a single $A$ player. When \eqref{eq:relation} holds, the number of $A$-player can be described by a one-dimensional Markov chain with absorbing state $0$ and $(1-p_S)\cdot N$.

Suppose that the number of $A$-players and $B$-players are $i$ and $(1-p_S)N - i$ respectively.

A $B$-player is chosen to be eliminated with probability $p_B$. The fitness of each $A$ neighbor of the focal $B$-player is given by:
\begin{equation*}
    f_A = 1 - \omega + \omega \cdot \left[b + (k-1)\cdot (q_{A|A}a + q_{B|A}b + q_{S|A}\eta)\right]
\end{equation*}
The fitness of each $B$ neighbor of the focal $B$-player is given by:
\begin{equation*}
    f_B = 1 - \omega + \omega \cdot \left[d + (k-1)\cdot (q_{A|B}c + q_{B|B}d + q_{S|B}\eta)\right]
\end{equation*}
Thus we have
\begin{equation*}
    T_A^+(i) = p_B \cdot \frac{kq_{A|B} \cdot f_A}{kq_{A|B} \cdot f_A + kq_{B|B} \cdot f_B}
\end{equation*}

\vspace{3ex}

A $A$-player is chosen to be eliminated with probability $p_A$. The fitness of each $A$ neighbor of the focal $A$-player is given by:
\begin{equation*}
    g_A = 1 - \omega + \omega \cdot \left[a + (k-1)\cdot (q_{A|A}a + q_{B|A}b + q_{S|A}\eta)\right]
\end{equation*}
The fitness of each $B$ neighbor of the focal $A$-player is given by:
\begin{equation*}
    g_B = 1 - \omega + \omega \cdot \left[c + (k-1)\cdot (q_{A|B}c + q_{B|B}d + q_{S|B}\eta)\right]
\end{equation*}
Thus we have
\begin{equation*}
    T_A^-(i) = p_A \cdot \frac{kq_{B|A} \cdot g_B}{kq_{A|A} \cdot g_A + kq_{B|A} \cdot g_B}
\end{equation*}

\vspace{3ex}

The fixation probability of a single $A$ player is given by\cite{nowak2004emergence}:
\begin{equation*}
    \rho_A = \phi_A(1/(1-p_S)N) = \frac{1}{1+\sum_{i=1}^{(1-p_S)N-1}\prod_{j=1}^i (T_A^-(j)/T_A^+(j))}
\end{equation*}

The ratio of the fixation probabilities of $A$ and $B$ is given by\cite{nowak2006evolutionary}:
\begin{equation*}
    \frac{\rho_A}{\rho_B} = \prod_{i=1}^{(1-p_S)N-1} \frac{T_A^+(i)}{T_A^-(i)},
\end{equation*}

\section{Results}

Note that
\begin{equation*}
    \begin{aligned}
        \prod_{j=1}^i \frac{T_A^-(j)}{T_A^+(j)} &= 1 + \omega \cdot\sum_{j=1}^i \left[q_{A|A}(G_B-G_A) - q_{B|B}(F_A-F_B)\right] + O(\omega^2). \\
    \end{aligned}
\end{equation*}
The terms with $\eta$ always have the same coefficients in $F_A$ and $F_B$, and they cancel each other out in $F_A - F_B$ and $G_A - G_B$. Thus the fixation probability is independent of the value of $\eta$.

\vspace{3ex}

Natural selection favors $A$, i.e., $\rho_A > 1/(1-p_S)N$\cite{nowak2004emergence} if and only if
\begin{equation*}
    \sum_{i=1}^{(1-p_S)N-1} \sum_{j=1}^i \left[q_{A|A}(G_A-G_B) - q_{B|B}(F_B-F_A)\right] > 0
\end{equation*}
which holds if and only if the following expression is positive:
\begin{equation*}
    \begin{aligned}
    &a \left(k^{2} + 2 k + 1\right) + b \left(2 k^{2} - 2 k - 1\right) + c \left(- k^{2} + k - 1\right) + d \left(- 2 k^{2} - k + 1\right) \\
    &+ p_{S}^{2} \left(a k^{2} - 2 a k + a + 2 b k^{2} - 4 b k + 2 b - c k^{2} + 2 c k - c - 2 d k^{2} + 4 d k - 2 d\right) \\
    &+ p_{S} \left(- 2 a k^{2} + 2 a - 4 b k^{2} + 6 b k - 2 b + 2 c k^{2} - 3 c k + c + 4 d k^{2} - 3 d k - d\right)
    \end{aligned}
\end{equation*}

\vspace{3ex}

Strategy $A$ is favored over strategy $B$, i.e., $\rho_A > \rho_B$, if and only if
\begin{equation*}
    \sum_{i=1}^{(1-p_S)N-1} \left[q_{A|A}(G_A-G_B) - q_{B|B}(F_B-F_A)\right] > 0
\end{equation*}
which holds if and only if the following expression is positive:
\begin{equation*}
    a \left(k + 1\right) + b \left(k - 1\right) + c \left(1 - k\right) + d \left(- k - 1\right) + p_{S} \left(k - 1\right) \left[c + d - a - b\right]
\end{equation*}

\vspace{3ex}

Applying the above results to the donation game, the payoff matrix is given by
\begin{equation*}
    \begin{pNiceMatrix}[first-row,first-col]
        & A & B \\
        A & b-c & -c \\
        B & b & 0 \\
    \end{pNiceMatrix}
\end{equation*}
where $b$ is the benefit of receiving a donation, and $c$ is the cost of making a donation, and $b>c>0$.

We have the following theorem:

\begin{theorem}
    In the donation game, using death-birth updating, in the limit of weak selection, the condition $\rho_A>1/(1-p_S)N$ and $\rho_A>\rho_B$ are equivalent, and they are both equivalent to the condition
    \begin{equation*}
        b/c > k - p_S(k-1).
    \end{equation*}
\end{theorem}

From the theorem, we see that the critical benefit-to-cost ratio $b/c$ decreases uniformly with the increse of $p_S$. As $p_S\to 1$, the critical $b/c$ value approaches $1$. The presence of the benevolence play two roles: (1) they offer a fixed payoff to the opponents, but this does not affect the fixation probability; (2) they reduce the actual degree of the active players, which leads to a decrease of the critical $b/c$ value.

\vspace{3ex}

\begin{figure}[H]
    \centering
    \includegraphics[width=1\textwidth]{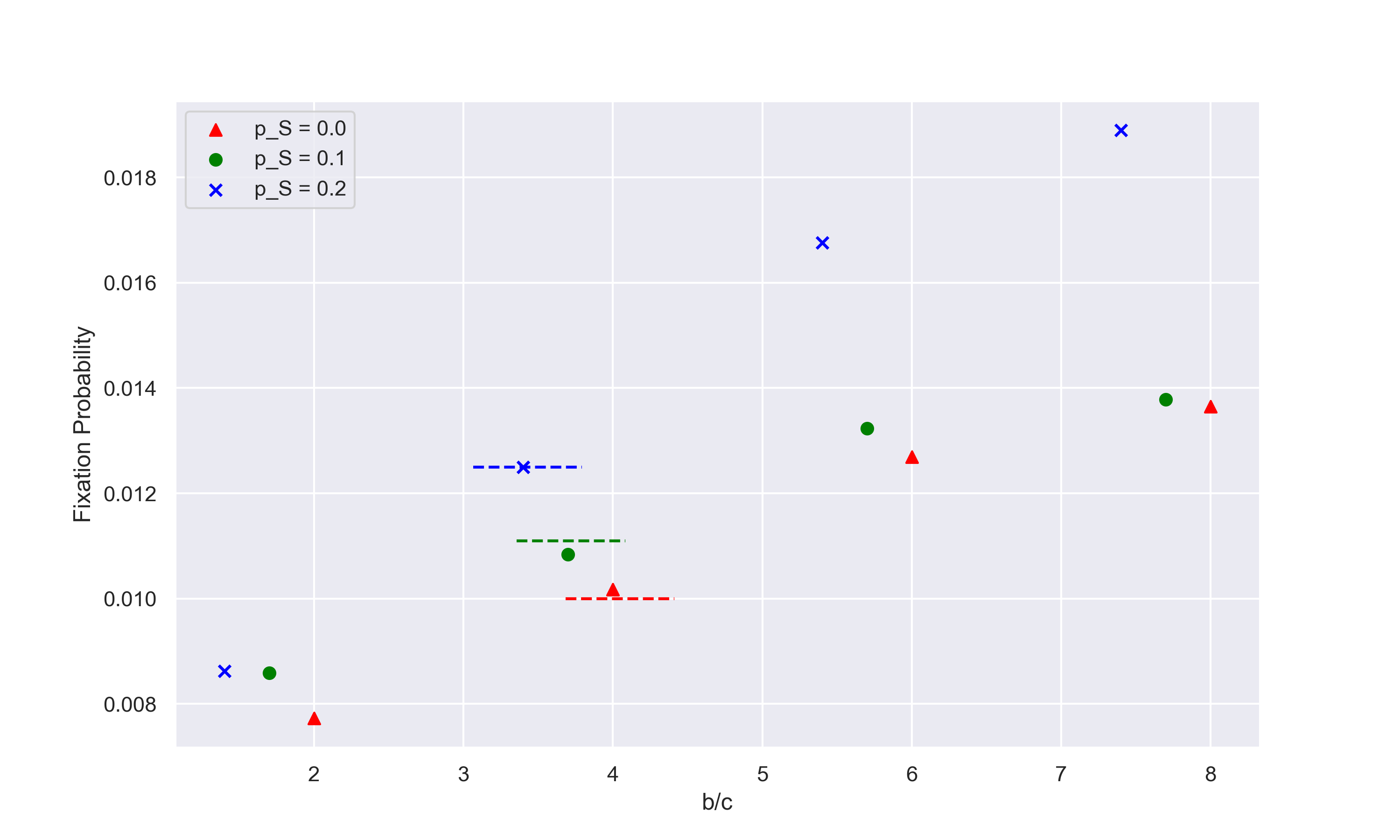}
    \caption{The fixation probability of a single cooperator in the donation game on a regular graph of degree $k=4$. The total population size is $N=100$, and the intensity of selection $\omega=0.01$. The dashed line is the theoretical prediction of the neutral fixation probability for each value of $p_S$. The $b/c$ value of the dots inside each dashed line is the critical value for the corresponding $p_S$ value. With the increase of $p_S$, the critical $b/c$ value decreases uniformly.}
    \label{fig:lattice_4}
\end{figure}

We perform numerical simulations to verify our results, as shown in Fig. \ref{fig:lattice_4}. The results of the numerical simulations are consistent with our theoretical predictions.

\bibliographystyle{unsrt}
\bibliography{ref}

\end{document}